\documentclass[lengthestimate,tighten,amssymb,prl,aps,twocolumn,floatfix,tightenlines]{revtex4}
\usepackage{graphicx,subfigure,dsfont}
\newcommand{\ket}[1]{\left | #1 \right \rangle}

\begin{document}
\title{Shor's quantum factoring algorithm on a photonic chip}
%\vspace{-0.3cm}}
\author{Alberto Politi}
\altaffiliation{These authors contributed to the work equally}
%\affiliation{Centre for Quantum Photonics, %H. H. Wills Physics Laboratory \& Department of Electrical and Electronic Engineering, University of Bristol, Merchant Venturers Building, Woodland Road, Bristol, BS8 1UB, UK\vspace{-1cm}}

%\affiliation{Centre for Quantum Photonics, H. H. Wills Physics Laboratory \& Department of Electrical and Electronic Engineering, University of Bristol, Merchant Venturers Building, Woodland Road, Bristol, BS8 1UB, UK}

\author{Jonathan C. F. Matthews}
\altaffiliation{These authors contributed to the work equally}
%\affiliation{Centre for Quantum Photonics, %H. H. Wills Physics Laboratory \& Department of Electrical and Electronic Engineering, University of Bristol, Merchant Venturers Building, Woodland Road, Bristol, BS8 1UB, UK}

\author{Jeremy L. O'Brien}
%\vspace{-0.3cm}}
\email{Jeremy.OBrien@bristol.ac.uk}
\affiliation{Centre for Quantum Photonics, %H. H. Wills Physics Laboratory \& Department of Electrical and Electronic Engineering, University of Bristol, 
Merchant Venturers Building, Woodland Road, Bristol, BS8 1UB, UK}
%\vspace{-0.7cm}}

%\date{\today}

\begin{abstract}
Shor's quantum factoring algorithm finds the prime factors of a large number exponentially faster than any other known method Ñ a task that lies at the heart of modern information security, particularly on the internet. This algorithm requires a quantum computer Ñ a device which harnesses the `massive parellism' afforded by quantum superposition and entanglement of quantum bits (or qubits). We report the demonstration of a compiled version of Shor's algorithm on an integrated waveguide silica-on-silicon chip that guides four single-photon qubits through the computation to factor 15. 
\end{abstract}

\maketitle

The realization of a quantum computer presents an exciting prospect of modern science. The processing of information encoded in quantum systems admitting quantum superposition and entanglement enables exponentially greater power for particular tasks. Originally conceived in the context of simulating complex quantum systems, it was the development of Shor's quantum factoring algorithm \cite{sh-conf-94-124} that showed the capability of factoring the product of two large prime numbers exponentially faster than any known conventional method \cite{shor-note-1} which has ignited efforts to fabricate such a device.

Despite progress towards this goal, proof-of-principle demonstrations of Shor's algorithm have so far only been possible with liquid-state NMR \cite{va-nat-414-883} and bulk optical implementations of simplified logic gates \cite{lu-prl-99-250504,la-prl-99-250505}, owing to the need for several logic gates operating on several qubits, even for small-scale compiled versions. However, these approaches cannot be scaled to a large number of qubits due to the purity, size and stability limitations of these systems. We demonstrate a compiled version of Shor's algorithm operating on four qubits in which the processing occurs in a photonic circuit of several one- and two-qubit gates fabricated from integrated optical waveguides on a silica-on-silicon chip \cite{po-sci-320-646,ma-natphot-3-346}. Whereas the full ShorÕs algorithm is designed to factorize any given input, a compiled version is designed to find the prime factors of a specific input.

The quantum circuit our device implements is the compiled version of Shor's algorithm for factorizing 15 \cite{va-nat-414-883,lu-prl-99-250504,la-prl-99-250505} (Fig 1A). This algorithm uses five qubits, one of which, $x_0$, is effectively redundant as it remains in a separable state throughout. The physical implementation (Fig. 1B) consists of two non-deterministic controlled-phase (CZ) gates (each with success $P = 1/9$, conditional on post selection) and six one-qubit Hadamard (H) gates \cite{ob-sci-318-1567}. The computation proceeds as follows (Figs. 1A, B): Four photons are input into the ``0'' or ``1'' waveguides to prepare the initial state $|\psi_{in}\rangle=|0\rangle_{x_1}|0\rangle_{x_2}|0\rangle_{f_1}|1\rangle_{f_2}$ (this does not represent 15, but rather the initialization for the compiled algorithm to compute the factors of 15). The H gates, implemented by 1/2 reflectivity directional couplers, then prepare each qubit in a superposition of 0 and 1, such that the entire state is a superposition of all possible four-bit inputsÑpart of the `massive parallelism' that gives rise to quantum speed-up. The core process is then performed by two independent $CZ$ gates, each implemented by a network of three $1/3$ directional couplers, that create a highly entangled output state \cite{lu-prl-99-250504,la-prl-99-250505}. Measurement of the output state of qubits $x_1$ and $x_2$, and classical processing, gives the results of the computation (see appendix). %\cite{shor-note-3}.

\begin{figure}[t!]
\begin{center}%$
\includegraphics[width=0.5\textwidth]{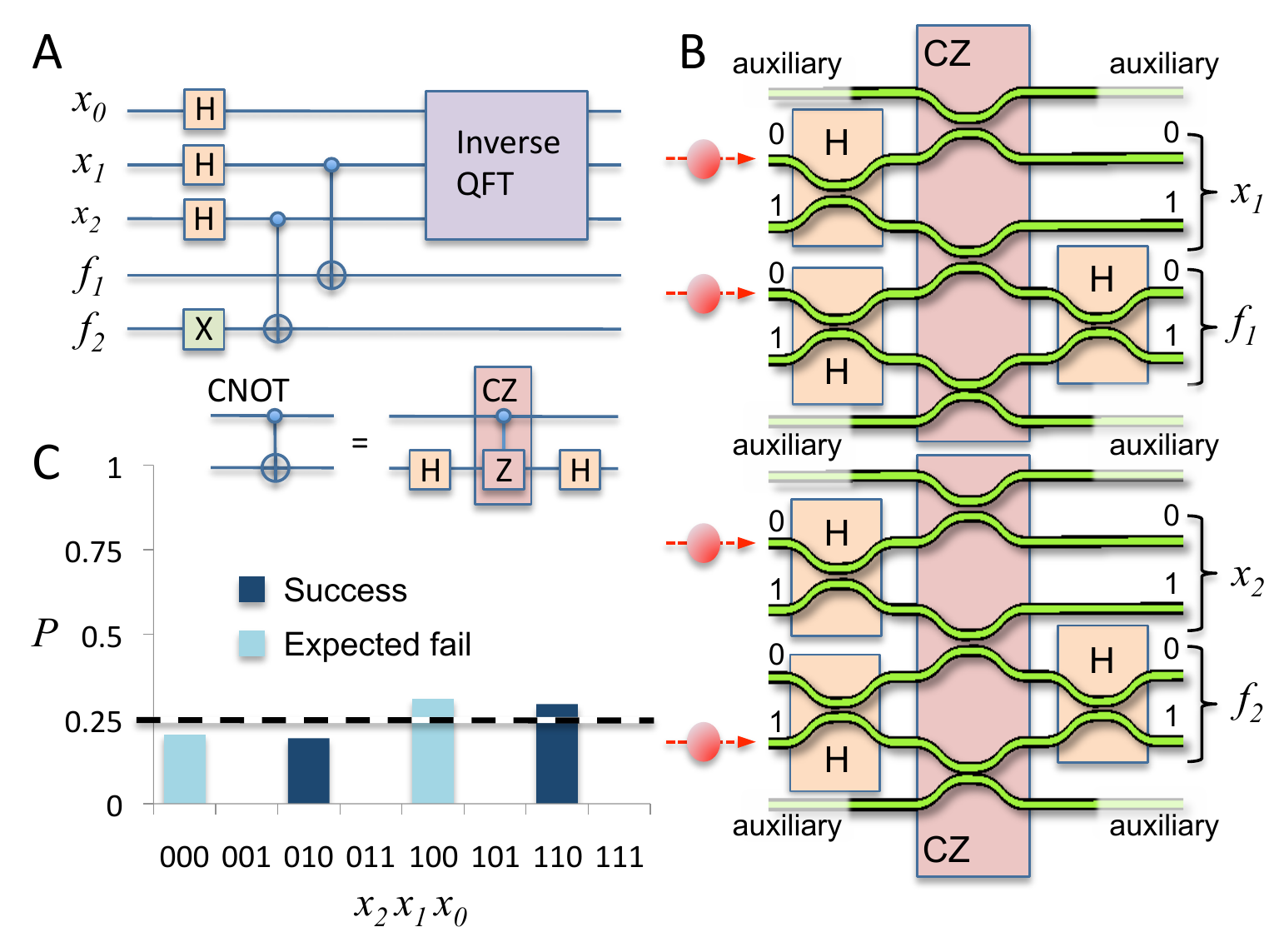}
\vspace{-1cm}
\end{center}
\caption{Integrated optical implementation of Shor's quantum factoring algorithm. (\textbf{A}) The quantum circuit. (\textbf{B}) Schematic of the waveguide on chip device that implements the quantum computation. The $x_n$ qubits carry the result of the algorithm; $f_n$ are additional qubit required for the computation to work.  (\textbf{C}) Outcomes of the algorithm.}
\vspace{-0.5cm}
\end{figure}

We simultaneously prepared four 790 nm photons via parametric down conversion, coupled them into and out of the chip with butt-coupled arrays of optical fibres, and detected them with silicon avalanche photo diodes at a typical coincidental rate of 100 Hz per measurement (integrated for 30 s). We input the state $\left|\psi_{in}\right\rangle$ and measured the output state of qubits $x_1$ and $x_2$; the output statistics (Fig. 1C) show the four binary outcomes $000$, $010$, $100$, $110$ (including the $x_0$ qubit). Outputs $010$ and $110$ lead to the correct calculation for finding the order $r = 4$ for the algorithm (see appendix), 
%\cite{shor-note-3}
which then enables efficient classical computation of the factors 3 and 5; 100 gives the trivial factors (1 and 15); and 000 is an expected failure mode inherent to Shor's algorithm. The measured results have a fidelity of $99 \pm 1\%$ with the ideal probability distribution (dashed line).

This demonstration of a small-scale compiled Shor's algorithm on a chip shows promise for quantum computing in integrated waveguides. While it currently uses an inefficient single photon source and modest efficiency detectors, ongoing progress to address heralded gates, efficient sources and detectors \cite{ob-sci-318-1567} combined with the results presented here will allow large-scale quantum circuits on many qubits to be implemented. Any quantum computer is a many-particle, many-path interferometer; the capability to implement such complex interferometers in a stable and miniaturized architecture is therefore critical to the future realization of large-scale quantum algorithms.

\section{Appendix}

The device in Fig. 1B performs a compiled version of the quantum routine in Shor's algorithm.
Euler's theorem dictates that for all coprime integers $a$ and $N$, there exists a least power $r$ known as the order of $a$ modulo $N$, such that both $a^r \equiv 1 \textrm{ mod } N$ and $1\leq r<N$. Provided $r$ is even, then it follows that $a^r-1 = \left(a^{r/2} -1\right)\left(a^{r/2}+1\right) \equiv 0 \textrm{ mod } N$, implying $N$ divides the product $\left(a^{r/2} -1\right)\left(a^{r/2}+1\right)$. Since $r$ is the order of $a$ modulo $N$, and provided  $a^{r/2} \neq-1 \textrm{ mod } N$, then it follows that the factors of $N$ must each divide $\left(a^{r/2} -1\right)$ and $\left(a^{r/2} +1\right)$. This therefore implies the factors of $N$ are given by the greatest common divisors of $N$ and $\left(a^{r/2} \pm1\right)$ (computed efficiently using Euclid's classical algorithm).  The most challenging part of Shor's algorithm, requiring the power of quantum computation, is to therefore find the order $r$ of some randomly chosen coprime $a$ of $N$. This is achieved by using entanglement and superposition across two registers of qubits (the argument $x_i$ and function $f_i$ registers) to compute the modular exponential function (MEF) $a^z\textrm{ mod } N$; the resulting interference in a subsequent quantum Fourier transform (QFT) yields high peaks in amplitude for the argument qubits, from which $r$ is determined (equivalent to the period of the MEF).

The compiled version of Shor's algorithm we used computes the prime factor of 15 when $a=2$ is chosen. This compilation %we used %follow, as described in Ref. $[5]$ of the main text, 
reduces the required number of function qubits ($f_i$) by evaluating $\textrm{log}_a[a^z\textrm{ mod } N]$ in place of $a^zx\textrm{ mod } N$; reducing the number of function qubits from $\textrm{log}_a[N]$ to $\textrm{log}_2[\textrm{log}_a[N]]$. The Inverse QFT shown in Fig. 1A does not need to be implemented in the quantum circuit for this case (nor for any $r=2^l, l \in \mathds{N}$), and can be performed by classical processing. %on the resultant measured qubits; this further simplification achieves the computation with the circuit shown in Fig 1B. 
Part of this classical processing inverts the order of the computed quantum bits, as in Fig 1C.

%The performance of  two two-qubit circuits, consisting of one $CZ$ and three $H$, was quantified by the truth table shown in figure Fig. 1C; the two two-qubit circuits have similarities with the ideal process of $88\pm1\%$ and $89\pm1\%$.  These values are limited by the photon source and coupler ratios; values as high as $93\pm1\%$ have otherwise be obtained.

%We characterized the two two-qubit circuits, consisting of one $CZ$ and three $H$, used for the computation by measuring the truth tables obtained by inputting and measuring qubits in all of the computational basis states (the algorithm requires just one input state). We measured similarities with the ideal process of $S=88\pm1\%$ and $S=89\pm1\%$.  These values are limited by the photon source and coupling ratios; values as high as $93\pm1\%$ have otherwise been obtained. The fidelity of the output probability distribution (Fig. 1C) with the ideal was $99\pm1\%$.

On inputting the state $\ket{\psi_{in}}=\ket{0}_{x1}\ket{0}_{x2}\ket{0}_{f1}\ket{1}_{f2}$ the circuit is designed to produce a product of two maximally entangled Bell pairs: $\frac{1}{2}(\ket{0}_{x1}\ket{0}_{f1}+\ket{1}_{x1}\ket{1}_{f1})(\ket{0}_{x2}\ket{1}_{f2}+\ket{1}_{x2}\ket{0}_{f2})$. Tracing out the $f_i$ qubits allows the outputs of the quantum computation to be obtained by measuring the $x_i$ qubits in the computational basis, which combined with the `redundant' qubit $x_0$ in the zero state, yields the 3-bit output: $x_2,x_1,x_0=000$, 010, 100, or 110. These 3-bit numbers correspond to the values $0, 2, 4$ and $6$, respectively. The first number is an expected failure inherent to Shor's algorithm, while the third yields the trivial factors 1 and 15. The second and fourth outcomes allow the calculation of the order $r=4$, which via Euclid's classical algorithm efficiently yields the correct prime factors 3 and 5.  This routine by itself has a success rate of 1/2; on repeating $n$ times it yields an overall success rate of $1-(1/2)^n$.
%This routine has a success rate of $1/2$, such that repeating the algorithm $n$ times yields an overall success rate of $1-(1/2)^n$, the standard result for Shor's algorithm.

Each single qubit Hadamard operation is deterministically realized using a 1/2 reflectivity directional coupler. Each non-deterministic controlled phase gate (CZ) consists of three 1/3 reflectivity directional couplers: quantum interference at the central coupler imparts the controlled phase shift on the $f_i$ qubit conditional on the $x_i$ qubit being in the $\ket{1}_{x_i}$ state; the other two 1/3 couplers balance the gate success probability (1/9), conditional on detecting one photon in the $x_i$ modes and one photon in the $f_i$ modes. 

%\begin{figure}[h]
%\begin{center}%$
%\includegraphics[width=0.5\textwidth]{ChipPhoto3.JPG}
%\vspace{-0.5cm}
%\end{center}
%\caption{The photonic waveguide chip used in our demonstration.}
%\end{figure}

\begin{acknowledgements}
We thank R. Jozsa, A. Laing, A. Montanaro, S. Takeuchi and X. Q. Zhou for helpful discussions.
\end{acknowledgements}

\end{document}